\documentclass[twocolumn]{revtex4}
\usepackage{amssymb}
\usepackage{amsmath}
\usepackage{graphicx}
\usepackage{epsfig}

\begin{document}

\title{\bf Specific-heat measurements of superconducting NbS$_{2}$ single crystal in an external magnetic field: Study on  the energy gap structure}

\author{J. Ka\v cmar\v c\'ik$^{1},$ Z. Pribulov\'a$^{1}$, C. Marcenat$^{1,2}$, T. Klein$^{3,4}$, P. Rodi\`ere$^{3}$, L. Cario$^{5}$, and P. Samuely$^{1}$ }

\affiliation{$^1$Centre of  Very Low Temperature Physics at
Institute of Experimental Physics, Slovak Academy of Sciences,
Watsonova
47, SK-04001 Ko\v sice, Slovakia\\
$^2$ CEA-Institut Nanosciences et Cryog\'enie/ UJF-Grenoble 1, SPSMS, UMR-E 9001, LaTEQS, 17 rue des martyrs, 38054 GRENOBLE, France\\
$^3$Institut N\'eel, CNRS, 38042 Grenoble, France
$^4$Institut Universitaire de France and Universit\'e Joseph Fourier, 38041 Grenoble, France\\
$^5$Institut des Mat\'eriaux Jean Rouxel, Universit\'e de
Nantes-CNRS, 44322 Nantes, France}

\date{\today}

\begin{abstract}
The heat capacity of  a  2H-NbS$_2$ single crystal has been
measured  by  a highly sensitive ac technique down to 0.6 K and in
magnetic fields up to 14 T. At very low temperatures data show
excitations over an energy gap ($2\Delta_S/k_BT_c \approx 2.1$)
much smaller than the BCS value. The overall temperature
dependence of the electronic specific heat $C_e$ can be explained
either by the existence of a strongly anisotropic single-energy
gap or within a two-gap scenario with the large gap about twice
bigger than the small one. The field dependence of the Sommerfeld
coefficient $\gamma$  shows a strong curvature for both principal-field orientations, parallel ($H||c$) and perpendicular ($H\perp
c$) to the $c$ axis of the crystal, resulting in a magnetic field
dependence of the superconducting anisotropy. These features are
discussed in comparison to the case of MgB$_2$ and to the data
obtained by scanning-tunneling spectroscopy. We conclude that the
two-gap scenario better describes the gap structure of NbS$_2$
than the anisotropic $s$-wave model.
\end{abstract}

\pacs{PACS     numbers: 74.50.+r, 74.7o.Dd} PACS: 74.25.Bt
74.25.Ha  74.70.Ad \maketitle


\section{INTRODUCTION}

The old concept of multiband/multigap superconductivity
\cite{suhl} has found its strong experimental evidence only
recently in the rich physics of magnesium diboride
\cite{mgdiboride}. Consequently more superconductors are
(re)considered along this line. One of the important examples are
the iron pnictides, a new class of high-$T_c$ superconductors
\cite{pnictides} for which multigap superconductivity is suggested
to lead to an exotic pairing mechanism with a sign reversal of the
order parameter between separated Fermi-surface sheets. A revision
in dichalcogenides brings more and more signatures of a
distribution of superconducting energy gaps, which can be either
due to different gaps on different Fermi-surface sheets or
anisotropic single gap.

Transition-metal dichalcogenides 2H-MX$_{2}$ (M = Nb, Ta, X = S,
Se) are materials with layered structure. Nb or Ta atoms are
trigonally prismatic coordinated by chalcogen atoms and metallic
layers are held together by weak van der Waals forces.
Because of this layered structure, electrical, magnetic, and
optical properties show a high degree of anisotropy. NbS$_{2}$ is
the only member of the 2H-MX$_{2}$ family, which does not undergo
a charge-density wave transition \cite{Moncton}. This could be a
reason for its strong anisotropy, much larger than that of
NbSe$_2$.

NbSe$_{2}$ had been considered for a long time as being a
conventional type II superconductor \cite{Matthias}. Later on,
effects of the anisotropic and strong coupling interactions were
taken into account \cite{Kobayashi, Sanchez}. Recent measurements sensitive
to the order parameter show evidence that more than one energy
scale is necessary to account for establishing superconductivity
\cite{Yokoya,Fletcher,Huang,Kiss,Vala}.  NbS$_{2}$ was also originally
considered as just another anisotropic superconductor and its
unusual specific heat dependence was not interpreted in  detail
\cite{Hamaue,Dahn}. An important breakthrough came with scanning
tunneling microscopy/spectroscopy (STM) measurements by
Guillam\'on et al. \cite{Guillamon}, showing  strong indications
for two superconducting energy gaps instead of a single
anisotropic one. Since STM is a surface probe, this strong
statement certainly needs independent support showing that the two
gaps are reflecting the bulk properties of the system. In this
paper we address this issue with bulk thermodynamic measurements
of the specific heat at temperatures down to 0.6 K and in magnetic
fields up to 14 Tesla via ac-calorimetry technique.  We find that
the electronic specific heat $C_e$ cannot be described by the
standard BCS model with a single isotropic energy gap. First, at
the lowest temperatures the data shows that quasiparticles are
excited over an energy gap much smaller than the BCS weak coupling
limit. The overall temperature dependence of $C_e$ can  be
described only if two gaps or an anisotropic one gap case is
considered. Second, the field dependence of the Sommerfeld
coefficient $\gamma$ shows a strong curvature in striking
similarity with that of NbSe$_2$ and MgB$_2$. However, the
anisotropy of $\gamma$ decreases with magnetic field in an
opposite manner compared to the latter system. Finally, the two
gap scenario is supported by the absence of in-plane gap
anisotropy in recent STM imaging of the vortex lattice in NbS$_2$
\cite{Guillamon}, and by the fact that  the numerical values of
the two gaps  obtained from fitting our data,
$2\Delta_S/k_BT_c\approx 2.1$ and $2\Delta_L/k_BT_c \approx 4.6$,
are also in a very good agreement with the STM data.

\section{EXPERIMENT}

Details of the  synthesis of the single crystalline samples can be
found elsewhere \cite{Fisher}. The crystals used for the specific
heat measurements come from the same batch as those used in the
previous  STM studies \cite{Guillamon}. Their chemical composition
was checked using an energy dispersive X-ray spectroscopy (EDS).
Single crystals were also confirmed to be of 2H polytype by X-ray
diffraction measurements. In our experiment a thin crystal with a
well defined hexagonal shape and dimensions 500x500x30 $\mu$m was
chosen.

Specific heat measurements have been performed  using an ac
technique as described elsewhere \cite{Sullivan}. The high
sensitivity of this technique is not only very well adapted to
measure the specific heat of very small samples but also to carry
continuous measurements during temperature or magnetic field
sweeps. We were thus able to obtain the field dependence of the
electronic part of $C/T$ at $T \approx 0.6$ K which only differs
from its zero temperature limit,  the Sommerfeld coefficient
$\gamma$ , by about 2$\%$. Measurements were performed with the
magnetic field aligned along the two main crystallographic
orientations, i.e. parallel and perpendicular to the basal $ab$
plane of the sample. The temperature oscillations of the sample
were recorded by a thermocouple calibrated in magnetic field using
measurements on ultrapure silicon. We performed measurements at
temperatures down to 0.6 K and in magnetic fields up to 8 T in the
$^3$He and $^4$He refrigerators in Ko\v sice. Supplementary
measurements up to 14 Tesla and down to 2 K were performed in
Grenoble.

\section{RESULTS AND DISCUSSION}

Figure 1 displays the temperature dependence of the specific heat
of the sample (plus addenda) in selected magnetic fields up to 8 T
for $H||ab$ and up to 3 T for $H||c$. The thermodynamic
superconducting  transition temperature at zero field was
determined from the local entropy balance around the phase
transition giving $T_c=$ 6.05 K. The zero-field anomaly at the
transition is sharp ( $\Delta$$T$$_{c}$ $\sim$ 0.4 K) indicating
the high quality and homogeneity of the sample.  The position of
the  specific-heat jumps  are gradually shifted toward lower
temperatures for increasing magnetic field. Despite a significant
broadening at high fields, the anomaly remains well resolved at
all fields. A field of 3 Tesla applied along the c-axis was
sufficient to completely suppress superconductivity in all the
temperature range. On the other hand, 8 Tesla applied along the
ab-planes shifts the superconducting anomaly down to only about
3-4K underlying the strong anisotropy of this system.

\begin{figure}[t]
\begin{center}
\resizebox{0.43\textwidth}{!}{\includegraphics{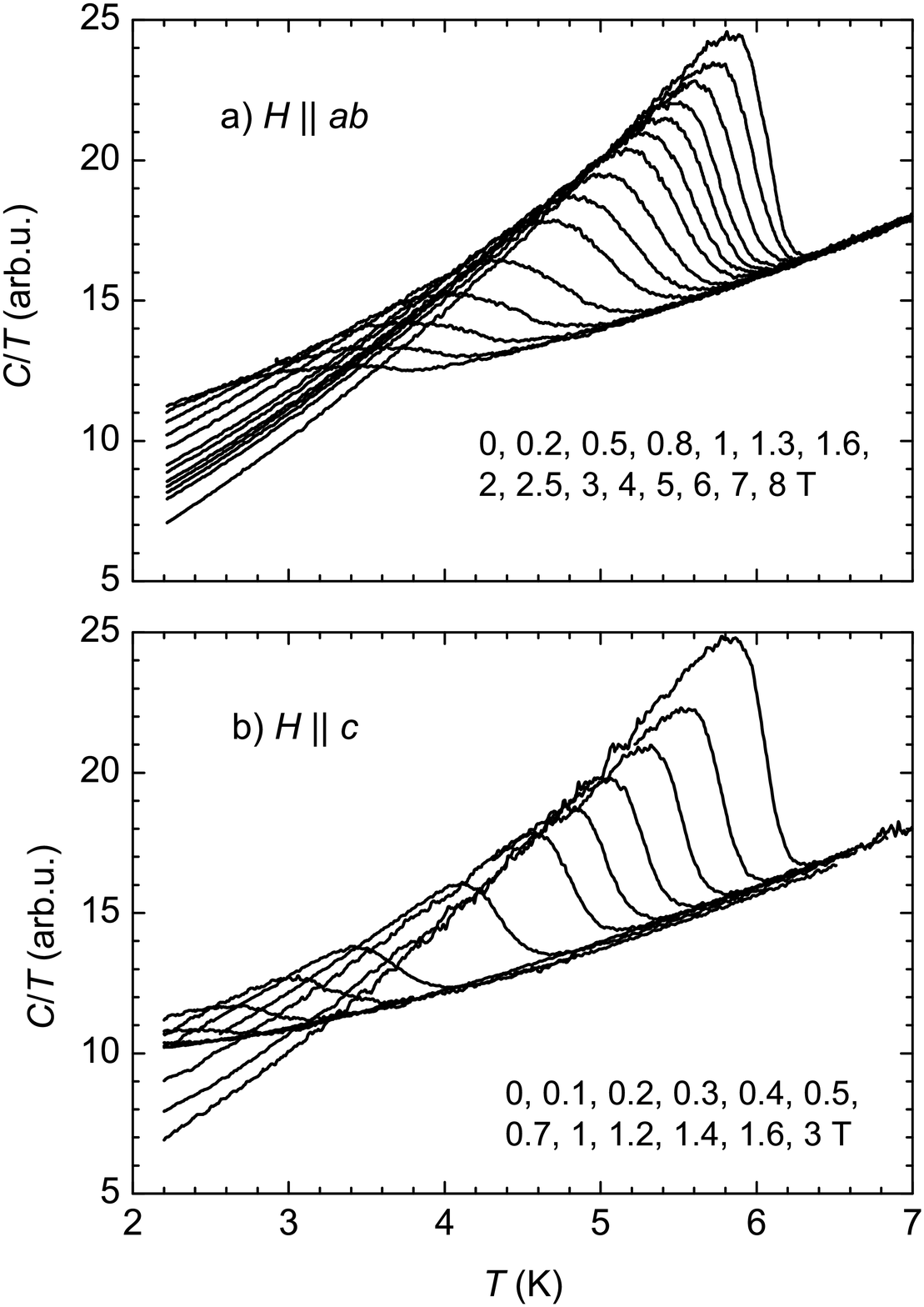}}
\caption{a) Total specific heat $C/T$ of NbS$_2$ in magnetic field
parallel (a) and perpendicular (b) to the $ab$ planes. In both
data sets the zero field measurement is the rightmost curve.}
\label{fig:fig1}
\end{center}
\end{figure}

Later we extended the measurements down to 0.6 K in a $^3$He
fridge where the specific heat was measured at zero field and at
$H||c=3$ T. In the case of a very small crystal like ours, it is
difficult to evaluate the exact total contribution of the addenda.
The electronic part of the total specific heat value can be
obtained by extrapolation of $C_{tot}/T$ for T approaching zero.
This value corresponds to $\sim 38\%$ of $\gamma_n$, with
$\gamma_n T$ being the electronic heat capacity of the sample in
the normal state. To avoid any fitting procedures, the addenda and
the phononic contributions have been eliminated by subtracting the
data taken at $H||c=3$ T from all the other runs. Thus, the
electronic specific heat of the sample, $C_e(T)$, normalized to
its normal state value, $\gamma_n T$, can be obtained
experimentally by: $\ C_e(T)/\gamma_nT = \frac {\Delta
(C/T)}{\gamma_n} +1$, where $\Delta (C/T)= (C(T,H=0)/T -
C(T,H=3$T$||c)/T$ and $\gamma_n$ =
$C(H=3$T$||c)/T_{|0.6K}-C(H=0)/T_{|0.6K}$. The result is presented
in Fig. 2 by the open circles. The only assumption in this
procedure is the absence of magnetic field dependence of the
addenda. This has been previously verified in numerous experiments
using the same thermocouple wires and also confirmed here
independently by the entropy conservation required for a second
order phase transition, proving the thermodynamic consistency of
the data and its treatment.

We first compare the electronic specific heat with the isotropic
single gap (ISG) BCS model. The dashed line in Fig. 2 presents the
ISG BCS specific heat (weak coupling of 2$\Delta/k_BT_{c} =
3.52$). One can see that while the height of the jump  at $T_c$ of
the experimental data is quite well reproduced, a significant
deviation occurs at lower temperatures. The discrepancy between
the BCS curve and the measured data represents 18 $\%$ (7 $ \%$)
of the total signal at 1.4K (4K), which is significantly larger
than the error bars of our measurements.

\begin{figure}[t]
\begin{center}
\resizebox{0.45\textwidth}{!}{\includegraphics{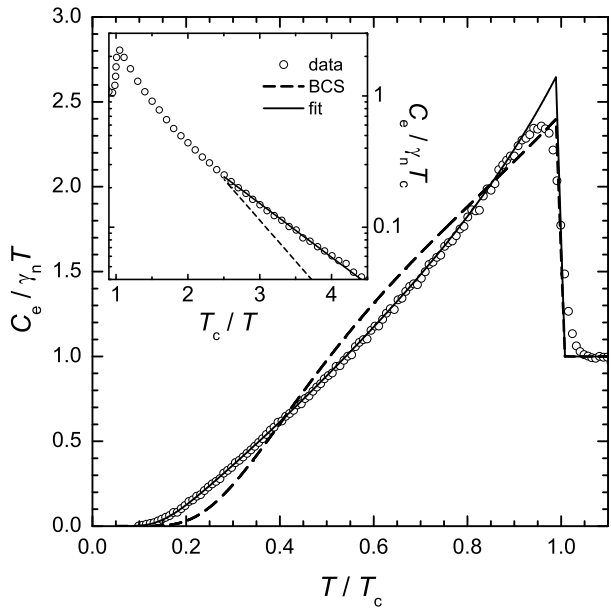}}
\caption{Open circles: Electronic specific heat of NbS$_2$ in zero
magnetic field extended down to 0.6 K. Dashed line: BCS single-gap
weak-coupling case. Solid line: two-gap model with
2$\Delta_S/k_BT_c$ = 2.1, 2$\Delta_L/k_BT_c$ = 4.6 and respective
relative contributions $\gamma_{S}/\gamma_{n}$ = 0.4,
$\gamma_{L}/\gamma_{n}$ = 0.6. The anisotropic gap model
with anisotropy parameter $\alpha$ = 0.5 and
2$\Delta_0$/k$_BT_{c}$ = 3.6 follows essentially the same line. Inset: exponential dependence of the
specific heat, the full line represents the best fit of the
exponential decay, the dashed line is the behavior expected for a
BCS single-gap weak-coupling limit.} \label{fig:fig2}
\end{center}
\end{figure}

The inset of Fig. 2 displays the logarithm of the electronic
specific heat versus $T_c/T$. As shown, one obtains an exponential
dependence $C_e \propto exp{(-b*T_c/T)}$ for $T_c/T \geq 2.5$.
However, the parameter $b$  is significantly lower than the value
expected for the BCS weak coupling limit in the temperature range
$T_c/T =2.5 - 4.5$ \cite{parks}. This corresponds to coupling
ratio 2$\Delta$/$k_BT_c \sim$ 2.3 that is much smaller than the
ISG BCS value of 3.52, indicating that the quasiparticles are
activated over a small energy gap. This fact as well as the
overall shape of the specific heat temperature dependence
resembles the case of MgB$_2$.

A phenomenological $\alpha$-model of the specific heat accounting
for independent contributions from two bands with two different
energy gaps has been successfully applied in the case of MgB$_{2}$
\cite{Bouquet}. The magnitude of the small gap 2$\Delta_S/k_BT_c$
and of the large gap 2$\Delta_L/k_BT_c$ at $T=0$ are fitting
parameters of the model. The third parameter is the relative
fraction of the density of states  of the  two bands
$\gamma_{S,n}/\gamma_{L,n}$. The full line in Fig. 2 represents a
fit to the experimental data yielding the following parameters:
2$\Delta_S/k_BT_c = $ 2.1 $\pm$0.05, 2$\Delta_L/k_BT_c =$
4.6$\pm$0.2, and $\gamma_{S,n}/\gamma_{L,n}$ = 0.67$\pm$ 0.15. The
value of the small gap is close to the one evaluated from the
exponential decay shown in the inset of Fig. 2. Importantly, both
gap values are in striking agreement with those found in the STM
experiment \cite{Guillamon} confirming that the latter are
characteristic of the bulk.

As previously shown by Huang et al. \cite{Huang}, the temperature
dependence of the specific heat of NbSe$_2$, another two-gap
superconductor, can also be described by an anisotropic $s$-wave
model, where the gap anisotropy is supposed to be in the form of
$\Delta=\Delta_0(1+\alpha\cos6\phi)$ corresponding to the
hexagonal in-plane symmetry. Here, $\Delta_0$ is the average gap
value and $\alpha$ denotes its anisotropy, yielding
$\Delta_{max}=\Delta_0(1+\alpha)$ and
$\Delta_{min}=\Delta_0(1-\alpha)$.  This model with parameters
$\alpha=0.5 $ and 2$\Delta_0/k_BT_c=3.6$ fits our data as well as
the two gap scenario, $the$ $difference$ $between$ $the$ $two$ $models$ $is$ $negligible$. We
remark that the anisotropic gap should leave its footprint in the
anisotropic vortex core  $\xi$ as it is proportional to the
related Fermi velocity $v_F$ divided by the gap at zero
temperature $\Delta(0)$. However, in contrast to NbSe$_2$, for
which STS images revealed a sixfold star shape of the vortex
cores, the fully isotropic vortices have been imaged in NbS$_2$
\cite{Guillamon} questioning the applicability of the anisotropic
single gap model in our case.

\begin{figure}[t]
\begin{center}
\resizebox{0.45\textwidth}{!}{\includegraphics{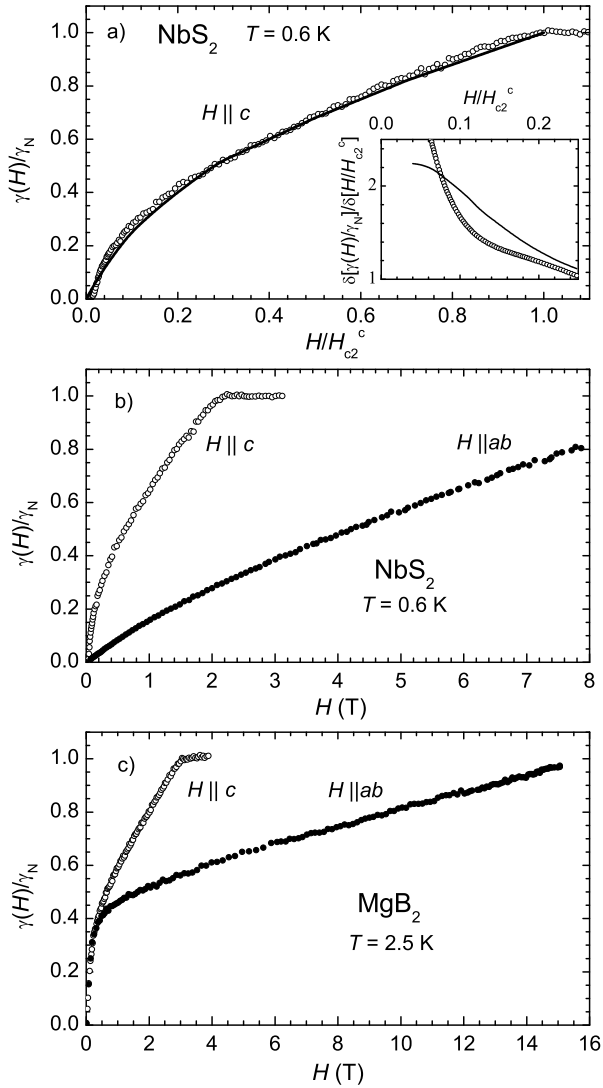}}
\caption{(a) Open circles - normalized Sommerfeld coefficient
$\gamma$ as a function of magnetic field perpendicular to the $ab$
planes of NbS$_2$. Line - model accounting for  highly
anisotropic gap with $\alpha$ = 0.5 \cite {Nakai}. Inset is the
derivative of the corresponding curves from the main panel: open
circles - of the measured data, line - of the model.  (b) and (c)
$\gamma/\gamma_N$ for both orientations of the magnetic field in
NbS$_2$ and MgB$_2$ \cite{Pribulova}, respectively.}
\end{center}
\end{figure}

 \begin{figure}[t]
\begin{center}
\resizebox{0.45\textwidth}{!}{\includegraphics{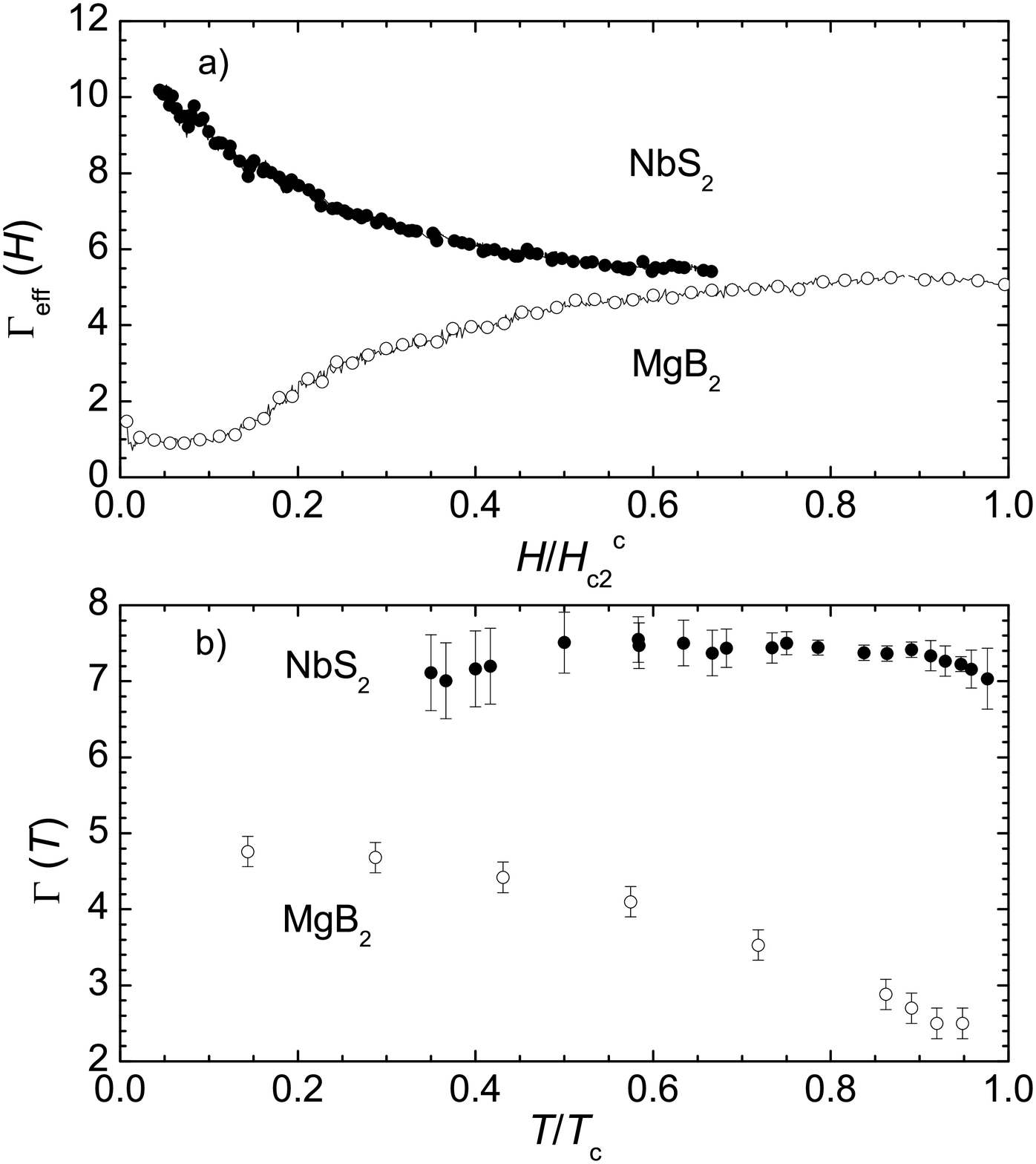}}
\caption{Anisotropy of NbS$_2$ (full circles) compared to MgB$_2$
\cite{Lyard,Pribulova} (open circles): (a) field dependence of
effective anisotropy defined as the ratio of the fields applied in
both principal orientations that correspond to the same $\gamma$
value in Fig.3b and Fig. 3c (b) temperature dependence of anisotropy
$\Gamma$ = $H_{c2}^{ab}/H_{c2}^c$.} \label{fig:fig5}
\end{center}
\end{figure}

We have measured thoroughly the evolution of the specific heat in
the mixed state. At $T=0.6$ K the electronic specific heat term
$C_e/T$ is very close to the Sommerfeld coefficient $\gamma$. Its
field dependence is displayed in Fig. 3a and 3b for  both
principal field orientations. Our maximum field available at this
temperature range (8 Tesla) was not sufficient to reach the normal
state for $H||ab$, but it was well above the upper critical field
value of 2.4$\pm$0.1 Tesla for $H||c$.   Fig. 3a emphasizes the
strong non-linearity of $\gamma(H)$ when $H$ is applied
perpendicular to the $ab$ plane. Again, such a non-linearity could
be associated with the existence of 2 gaps or a single anisotropic
one.

 The increase of $\gamma$ with magnetic field is mainly due to the quasiparticle
contribution inside the vortex cores. In the case of
superconductor with a single isotropic gap, $\gamma$ should
increase linearly in small magnetic field and a small nonlinearity
in $\gamma(H)$ appears above the field where flux lines start
overlapping. According to the calculations of Nakai et al.
\cite{Nakai}, much stronger non-linearity of $\gamma(H)$  is
achieved in case of anisotropic-gap superconductors. The full line
in the Fig. 3a displays the field dependence of the normalized
density of states $N(B)/N_0$ (proportional to the Sommerfeld
coefficient) calculated by Nakai et al. for the anisotropic gap with $\alpha=0.5$.
The model qualitatively reproduces the behavior in our data (open
circles), but fails to describe the fast increase of $\gamma$(H)
at low fields. This is evident in the inset of Fig. 3a where the
derivative $\partial{N(B)}/\partial{B}$, i.e. the slope of $N(B)$
for $\alpha=0.5$ (line) and the slope of measured Sommerfeld
coefficient $\partial{\gamma(H)}/\partial{H}$ (open circles) are
compared.  This discrepancy makes the explanation of the specific
heat data by the anisotropic gap scenario rather inconsistent. The
observed $\gamma(H)$ behavior resembles the two-gap case of
MgB$_2$ (see Fig. 3c, left curve) \cite{Pribulova}, where the initial rapid increase
of $\gamma(H)$ due to dominant role of the $\pi$ band with  the
small gap changes at fields where the $\sigma$ band with the large gap comes into
play.

\begin{figure}[t]
\begin{center}
\resizebox{0.45\textwidth}{!}{\includegraphics{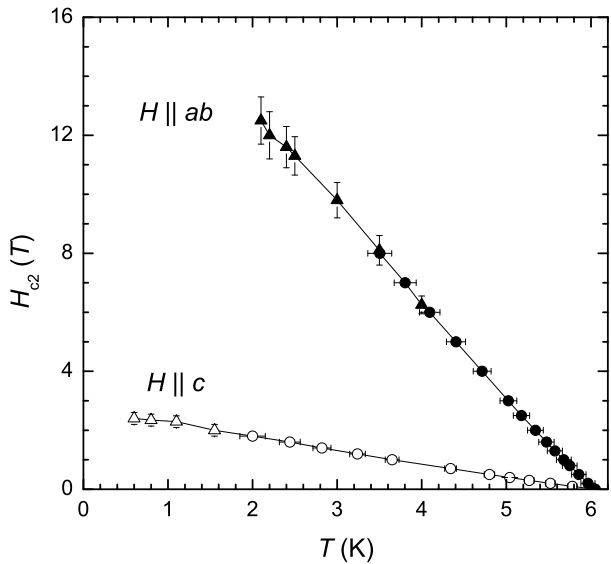}}
\caption{Upper critical field $H$$_{c2}$ for magnetic field $||ab$
(full circles result from temperature-sweep measurements, full
triangles from field sweeps) and $||c$ (open circles result from
temperature-sweep measurements, open triangles from field
sweeps).} \label{fig:fig4}
\end{center}
\end{figure}

Next, we compare the behavior of the Sommerfeld coefficients in
magnetic fields applied parallel and perpendicular to the basal
planes of the both materials,  NbS$_2$ (Fig. 3b) and MgB$_2$ (Fig.
3c). The latter is taken from Ref.\cite{Pribulova}. 
Following the
procedure introduced by Bouquet et al. in Ref.\cite{bouquet2} an
effective anisotropy $\Gamma_{eff}$ can be obtained from these $\gamma(H)$ dependences. 
It is defined as the ratio of the magnetic fields in the $ab$ plane and along the $c$ axis yielding the same $\gamma$ value in Fig. 3b and Fig. 3c. $\Gamma_{eff}$ is plotted in Fig 4a for both compounds as a function of the field $H||c$ normalized to its upper critical field value $H_{c2}||c$. 
As discussed  already in the work of Bouquet et al. on MgB$_2$ \cite{bouquet2} the choice of the abscissa is arbitrary. We chose $H||c$, but we could  plot $\Gamma_{eff}$ versus $H||ab$ or versus $\gamma(H)$ as well. 
Note that this
$\Gamma_{eff}$ tends  towards the usual anisotropy of H$_{c2}$,
$\Gamma = \frac {H_{c2}^{ab}}{H_{c2}^{c}}$ when
$\frac{\gamma}{\gamma_n} \rightarrow 1$ at large magnetic fields.
In MgB$_2$ at low fields ($\frac {H}{H_{c2}^{c}} < 0.1$), the $\gamma(H)$
curves for the two principal directions are practically identical which gives
$\Gamma_{eff} = 1$ as shown in Fig. 4a. At larger
fields, $\Gamma_{eff}$ increases reflecting a reduced contribution
from the isotropic $\pi$-band, reaching $\Gamma_{eff} \sim 5$
which is the anisotropy of the dominant $\sigma$-band
\cite{Pribulova}. In NbS$_2$, one observes an opposite field
dependence of $\Gamma_{eff}$ which starts from a highly
anisotropic value $\Gamma_{eff} \sim 10$ at low fields and
decreases to $\Gamma_{eff} \sim 5.5$ at our maximum field. A field
dependent superconducting anisotropy is a typical signature of
multigap superconductivity where a role of bands with different
gaps can significantly vary with magnetic field \cite{kogan,dahm,
golubov}. In contrast to MgB$_2$ case, in NbS$_2$ both bands would
be anisotropic, as suggested by analogy with NbSe$_2$ in which
band structure calculations  \cite{johannes} show mostly  4 Fermi
surface sheets derived from Nb $d$-bands forming warped cylinders
along the $c$ axis, centered on the $\Gamma$ and $K$ points in the
Brillouin zone. Moreover, two sheets derived from the bonding Nb
$d$ band are significantly more warped than the two derived  from
the antibonding  Nb $d$ band. Different warping of Nb sheets can
naturally lead to a different level of anisotropy in each band.  Thus, a qualitatively different behavior of $\Gamma_{eff}(H)$ compared
to MgB$_2$ can be expected.

Finally, we inspected the upper critical magnetic fields for both
principal orientations of magnetic field. Fig. 5 summarizes the
values of $H$$_{c2}$ derived from the temperature-sweep
measurements of the specific heat shown in Fig. 1, as well as from
field-sweep measurements. Two sets of  field-sweeps were
performed, one in 14 Tesla magnet for $H||ab$ in a temperature
range down to 2 K, and another one taken in the $^3He$ cryostat
down to 0.6 K in the 8 Tesla coil. As stated above, we determined
$T_c$ at zero field from the local entropy balance around the
anomaly. However, at finite fields this definition is not very
practical for establishing $T_{c2}(H)$, or $H_{c2}(T)$. In order
to reduce the uncertainty of the $H_{c2}$ value arising from the
broadening of the transition particularly at lower temperatures
(higher fields), we inspected the temperature shift between two neighboring curves in Fig. 1. A similar procedure was used
to determine $H_{c2}$ from field-sweep measurements. The resulting
temperature dependence of $H_{c2}$  is shown in Fig. 5 for both
$H||{ab}$ and $H||c$.

Importantly, the three independent sets of temperature and field-sweeps measurements yield consistent results with a nice overlap.
Both temperature dependencies show a slight positive curvature for
temperatures $T >T_c / 2$. The upper critical field in the $ab$
plane reveals very high values with $dH_{c2}/dT$ slope of about 3
Tesla/K, close to the Pauli paramagnetic limit.

The temperature dependence of the superconducting anisotropy
$\Gamma$ calculated as a ratio $H$$_{c2}||ab/H_{c2}||c$ of the
upper critical fields is displayed in Fig. 4b together with the
results obtained in MgB$_2$ \cite{Lyard}. As shown, in contrast to
MgB$_2$, the resulting anisotropy $\Gamma$ is  close to 7 and
approximatively constant for $T / T_c > 0.3$. Note that this value
might be slightly underestimated in case of a small misalignement
of the crystal for $H || ab$. Our results are consistent with
those obtained  by Onabe et al. \cite{Onabe} from resistive
measurements in a field up to 2 Tesla. The strong decrease of
$\Gamma$ in MgB$_2$ close to $T_c$ is a direct consequence of the
existence of the isotropic $\pi$- band. This is not a general
feature of multiband superconductivity since $\Gamma (T)$ results
from a subtle balance between the Fermi velocities and the
relative weight in the DOS of the different bands \cite{dahm}.
These precise calculations are still to be carried out in the case
of NbS$_2$.

\section{CONCLUSIONS}

Analysis of the zero-field specific heat data has shown that: (1)
the zero-field electronic term of the specific heat cannot be
described by an isotropic single-gap BCS formula, but it is
compatible with the two-gap $\alpha$ model; (2) the large (small)
gap is 2$\Delta_L/k_BT_c \approx$ 4.6 (2$\Delta_S/k_BT_c \approx$
2.1). The measurements in the mixed state have supported the
two-gap scenario revealing: (3) a strongly non-linear $\gamma
(H)$; (4) a field-dependent superconducting anisotropy.  Even if
some of these features of the specific heat could be eventually
explained by an extremely anisotropic-gap superconducting model,
this would not be compatible with the observation of two
well-resolved gap features with sizes of 2$\Delta_S/k_BT_c
\approx$ 2 and 4, respectively, and also with the absence of
in-plane anisotropy in the vortex lattice images by the scanning-tunneling spectroscopy of Guillam\'on et al.  To conclude, our
bulk thermodynamic measurements  are in full agreement with STM
spectra,  supporting that NbS$_2$ is another case of well resolved
two-gap superconductor.

\acknowledgments

This  work  was supported  by the EC Framework Programme
MTKD-CT-2005-030002, by the EU ERDF (European regional development
fund) grant No. ITMS26220120005, by the Slovak Research and
Development Agency, under Grants No. VVCE-0058-07,
No. APVV-0346-07, No. SK-FR-0024-09 and No. LPP-0101-06, and by the U.S. Steel Ko\v sice,
s.r.o.  Centre of Low Temperature Physics is operated as the
Centre of Excellence  of the Slovak Academy of Sciences. We thank G. Karapetrov for careful reading of the manuscript.

\end{document}